%% file: main.tex
\documentclass[journal]{IEEEtran}
\ifCLASSINFOpdf
\else
\fi

\input{packages}

\begin{document}
\title{Assessing Active Distribution Network Flexibility: On the Effects of Nonlinearities and Nonconvexities}

%
%
%

\author{Andrey~Churkin,
        Pierluigi~Mancarella,~\IEEEmembership{Senior Member, IEEE},
        and~Eduardo~A.~Martínez~Ceseña,~\IEEEmembership{Member,~IEEE}
\thanks{This work was carried out as a part of the ATTEST project (the Horizon 2020 research and innovation programme, grant agreement No 864298).}
\thanks{The authors are with the Department of Electrical and Electronic Engineering, the University of Manchester, UK (e-mails: \{andrey.churkin, p.mancarella, alex.martinezcesena\}@manchester.ac.uk). P. Mancarella is also with the Department of Electrical and Electronic Engineering, the University of Melbourne, Australia. E. A. Martinez Cesena is also with the Tyndall Centre for Climate Change Research, UK.}
}

%
%

\markboth{IEEE PES LETTERS}%
{Shell \MakeLowercase{\textit{et al.}}: Bare Demo of IEEEtran.cls for IEEE Journals}
%



\maketitle

\begin{abstract}
A widespread approach to characterise the aggregated flexibility of active distribution networks (ADNs) is to estimate the boundary of the feasible network operating areas using convex polygons in the P-Q space. However, such approximations can be inaccurate under realistic conditions where, for example, the nonlinear nature of the network is captured, and the behaviour of flexible units is constrained. This letter demonstrates, using a small ADN example with four flexible units and considering only nonlinearities from the network, that reaching the full P-Q flexibility areas would require perfect coordination of units and high ramping rates. Without these requirements, the P-Q areas become nonconvex. Thus, if the effects of nonlinearities and nonconvexities are ignored, existing approaches in the literature can result in overestimation of ADN flexibility and give rise to impractical solutions, hampering coordination between transmission and distribution system operators.
\end{abstract}


%
\IEEEpeerreviewmaketitle

\section{Introduction}
%
%
%
%



\IEEEPARstart{R}{ecent} advances in the integration of renewable and distributed energy resources (DERs), information and communications technologies, and network automation have led to the emergence of active distribution networks (ADNs). Such networks can aggregate and utilise the flexibility provided by flexible units - resources with the technical ability to control power exchange with the grid, e.g., controllable DERs, battery energy storage systems, prosumers, electric vehicles, etc. 
Recent studies, e.g., \cite{Silva2018,Capitanescu2018,Riaz2021}, have proposed characterising the flexibility of ADNs as a set of feasible network operating points in the P-Q space. This approach enables estimating the limits of aggregated flexible active and reactive power that an ADN can provide. Different methods have been used in the literature to approximate the boundary of the network feasibility area, including random sampling and optimal power flow (OPF) methods \cite{Bolfek2021,Contreras2021}. Such approximations are usually considered as a tool for coordinating transmission and distribution system operators (TSOs and DSOs) in flexibility procurement at the TSO/DSO interface.

However, despite the rapidly evolving research on ADN flexibility, existing studies \textbf{overlook nonlinearities and nonconvexities of network flexibility models} and do not consider related issues of ADN operation. This research gap arose due to the use of flexibility estimation algorithms that approximate the feasible network operating area in the P-Q space by convex polygons. Some studies, e.g. \cite{Silva2018,Riaz2021}, do not question the correctness of this convexification. Other works, e.g. \cite{Bolfek2021}, deliberately use linearised and convexified OPF formulations. 
Moreover, the focus has been placed on estimating the boundary of the network feasibility area. It is usually implied, for example in \cite{Stankovic2021}, that the larger this area approximated by the boundary, the more flexibility the network can provide, and the more useful TSO-DSO coordination can be achieved. 
By focusing on the boundaries of the flexibility area, existing studies inherently assume perfect DER coordination and active network management with fast unit response. In practice, however, such coordination will require incentives, information exchange, high ramping requirements, and other factors that may not be in place. Therefore, existing approaches in the literature can lead to overestimation of ADN flexibility and give rise to impractical solutions.


This letter aims to raise concerns about the nonlinearities and nonconvexities associated with ADN flexibility. The nonlinear behavior of flexible power provision is illustrated by the case study of a radial distribution network with four units. Only the nonlinearities of the network power flow model are considered, while other parameters are assumed linear (individual P-Q capabilities of flexible units and their cost functions) or omitted in the modelling (minimum stable level of generation for units, ramping constraints, etc.).
The results demonstrate that flexible units perform active network management to ensure the feasibility and cost-efficiency of certain operating points. For example, some units can produce flexible power while others consume it. This power swap effect enables alleviating voltage constraints to maximise the aggregated network flexibility.
The network feasibility set also contains areas where flexible units have to shift their power output rapidly and precisely between close operating points.
Such effects of nonlinearities and nonconvexities pose challenges to the operation of ADNs and should be accurately considered in future research on distribution network flexibility.



\section{Simulation Results and Discussion}\label{Section: results}
To illustrate the nonlinearities and nonconvexities of ADN flexible power provision, a cost-minimising OPF model was selected as a tool for network flexibility analysis. A generalised formulation of this model is given by \eqref{Flex_OPF: objective1}-\eqref{Flex_OPF: other constraints}. The objective function \eqref{Flex_OPF: objective1} minimises the cost of upward and downward regulation of flexible active and reactive power for flexible units located at each node of the system, $k \in \mathcal{K}$. Equations \eqref{Flex_OPF: p_delta} and \eqref{Flex_OPF: q_delta} define the volume of regulations depending on the initial flexible power output of units, $p_{k}^0$ and $q_{k}^0$, and the output at the new operating point, $p_{k}$ and $q_{k}$. The flexible power output of units is limited by the feasibility set $\mathcal{F}$ in \eqref{Flex_OPF: other constraints}, where the output of each unit is bounded by its P-Q capability, $\Delta\mathcal{S}_{k}$, and the network constraints are imposed depending on the selected OPF model, e.g., power balance and voltage limits for each node, and power flow limits for each line $l \in \mathcal{L}$. In this study, the analysis of network flexibility is based on the nonlinear DistFlow OPF formulation, which is an exact OPF model for radial distribution networks \cite{Baran1989-1,Baran1989-2}.

\begin{model}[t]
\caption{Cost-minimising flexibility estimation \hfill}
\label{Flex_OPF}
\begin{subequations} 
\vspace{-2\jot}
\begin{IEEEeqnarray}{llll}
    \IEEEeqnarraymulticol{3}{l}{ \min_{p_l,q_l,v_k,p^\uparrow_{k},p^\downarrow_{k},q^\uparrow_{k},q^\downarrow_{k}} \smashoperator{\sum_{k \in \mathcal{K}}} C^f_{k}(p^\uparrow_{k},p^\downarrow_{k},q^\uparrow_{k},q^\downarrow_{k})} \label{Flex_OPF: objective1} &\IEEEyesnumber\\
    \text{s.t.} & \IEEEnonumber\\
    p^\uparrow_{k} - p^\downarrow_{k} = p_{k} - p_{k}^0 \quad  &\forall k \in \mathcal{K} \qquad \label{Flex_OPF: p_delta} \\
    q^\uparrow_{k} - q^\downarrow_{k} = q_{k} - q_{k}^0 \quad  &\forall k \in \mathcal{K} \qquad \label{Flex_OPF: q_delta} \\
    p_{k},q_{k} \in \mathcal{F} = \Big\{\enskip\Delta\mathcal{S}_{k}: \qquad\qquad\qquad\qquad   &\forall k \in \mathcal{K} \label{Flex_OPF: other constraints} \\
    \qquad\qquad\underline{v_k} \leq v_k \leq \overline{v_k} \qquad\enskip \forall k \in \mathcal{K} & \IEEEnonumber\\
    \qquad\qquad{p_l}^2+{q_l}^2 \leq \overline{s_l}^2 \qquad \forall l \in \mathcal{L} \enskip \Big\} & \IEEEnonumber
    \vspace{-1\jot}
\end{IEEEeqnarray}
\end{subequations}
\end{model}

\begin{figure*}[t]
    \centering
    \includegraphics[width=\textwidth]{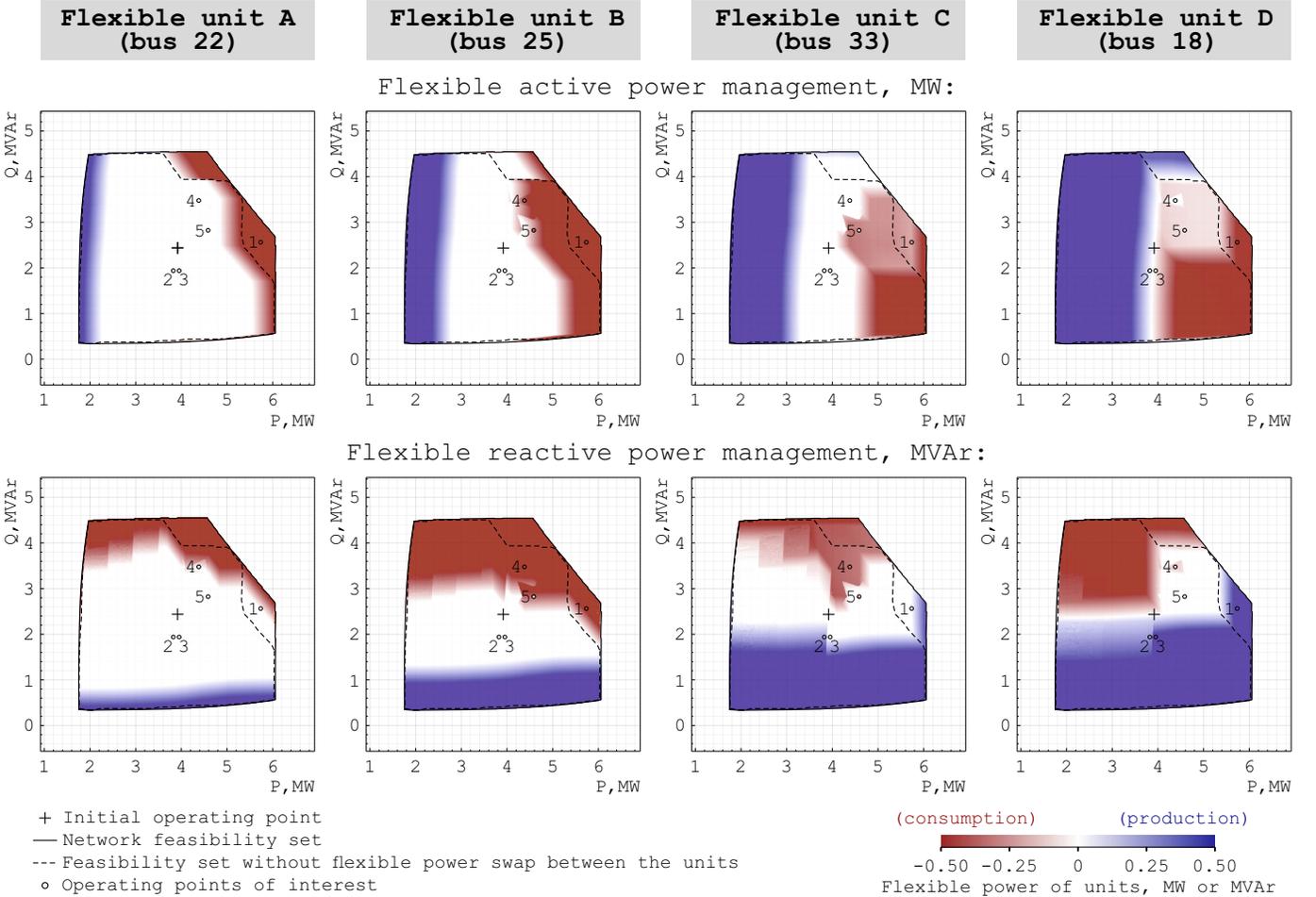}
    \caption{Flexible active and reactive power of the units for different flexibility requests defined by the cost-minimising OPF model, in MW and MVAr. The consumption and production of flexible power is differentiated using a red-blue heatmap. The areas characterise the aggregated network flexibility in the P-Q space at the TSO/DSO interface. The cross-markers correspond to the initial operating point, while the coordinates represent the network's power consumption.}
    \label{Fig: nonlinearities_PQ}
\end{figure*}

A popular case study, the IEEE 33-bus radial distribution network \cite{Baran1989-2}, is selected to illustrate the effects of nonlinearities and nonconvexities. Four flexible units with identical P-Q capabilities are placed in the network at buses 22, 25, 33, and 18 (one at the end of each feeder). Parameters of the units are specified in Table~\ref{table: data}. Note that the P-Q capability constraints and the cost functions are assumed to be linear. Thus, the only nonlinearities come from the network power flow model.
The buses with flexible units have different voltages, which imposes different constraints on the flexible power provision due to voltage limits. For example, units C and D operate at voltages close to the lower limit of 0.9 p.u. and therefore cannot increase their power consumption significantly.
Model \eqref{Flex_OPF: objective1}-\eqref{Flex_OPF: other constraints} was solved for multiple feasible operating points, both at the boundary of the feasible network operating area and within the boundary. Specifically, the model was solved 58,067 times for the discretised operating area with a step of 16.66 kVA. The results are displayed in Fig.~\ref{Fig: nonlinearities_PQ}, where the aggregated network flexibility is shown from the perspective of flexible power regulation by different units. Since there are four units regulating P and Q components of flexible power, the figure has a total of eight subplots. Moreover, the consumption and production of flexible power is differentiated using a red-blue heatmap. This visualisation enables a clear interpretation of the flexible unit actions at different feasible operating points.

\begin{table}[b]
\renewcommand{\arraystretch}{1.1}
\caption{Parameters of Flexible Units Placed in the Network}
\centering
\begin{tabular}{lllll}
\toprule 
\multirow{2}{*}{Parameter} & \multicolumn{4}{c}{Flexible unit}
\\
\cmidrule(l){2-5} 
& A & B & C & D\\ 
\midrule 
Bus \# & 22 & 25 & 33 & 18\\ 
Initial bus voltage (p.u.) & 0.991 & 0.970 & 0.917 & 0.913\\ 
P regulation limits (kW) & $\pm$500 & $\pm$500 & $\pm$500 & $\pm$500\\ 
Q regulation limits (kVAr) & $\pm$500 & $\pm$500 & $\pm$500 & $\pm$500\\
P cost (\$/kWh) & 0.375 & 0.350 & 0.325 & 0.300\\ 
Q cost (\$/kVArh) & 0.188 & 0.175 & 0.163 & 0.150\\
\bottomrule 
\end{tabular}
\label{table: data}
\end{table}




Since the operation of flexible units is limited by voltage constraints, the units cannot achieve their maximum total power consumption, which is displayed as the reduced upper right part of the network flexibility area.
Note that applying the cost-minimising OPF model \eqref{Flex_OPF: objective1}-\eqref{Flex_OPF: other constraints} for flexibility analysis results in solutions where cheaper flexible units get activated first, subject to the network constraints. It can be observed that units C and D (the cheaper ones) participate in the flexibility provision at operating points close to the initial operating point. Units A and B (the more expensive ones) get activated later, at the operating points closer to the feasibility boundary. However, the optimal operation of the flexible units exhibits highly nonlinear behavior. 
For example, at operating point \raisebox{.5pt}{\textcircled{\raisebox{-.9pt} {1}}}, flexible units B, C, and D consume active power and increase the total power consumption of the network.
Yet, unit D produces reactive flexible power. This counterintuitive solution is justified by the fact that unit D is located at the part of the network with a low voltage profile and cannot increase its power consumption further. Thus, unit D still produces reactive power, while unit B consumes it. This flexible power exchange (swap) enables units to alleviate the voltage constraints and maximise the network feasibility area. But, perfect coordination of units is required to work at such operating points, which might not be the case in realistic distribution networks. Without the flexible power swap, the network feasible operating area reduces and becomes nonconvex, as shown by the dashed boundary.

Another effect of the model's nonlinearities is the rapid shifts in flexible power output of the units between close operating points. This can be demonstrated by comparing flexible reactive power management at operating points \raisebox{.5pt}{\textcircled{\raisebox{-.9pt} {2}}} and \raisebox{.5pt}{\textcircled{\raisebox{-.9pt} {3}}}. At point \raisebox{.5pt}{\textcircled{\raisebox{-.9pt} {3}}}, the cheapest unit D is activated to produce reactive power. But, at the nearby point \raisebox{.5pt}{\textcircled{\raisebox{-.9pt} {2}}}, most of the reactive power production is shifted to unit C. This shift is justified by the cost-minimising model that identifies the least-cost flexible power production, subject to network constraints. Other notable examples of rapid and nonlinear active network management by flexible units are operating points \raisebox{.5pt}{\textcircled{\raisebox{-.9pt} {4}}} and \raisebox{.5pt}{\textcircled{\raisebox{-.9pt} {5}}}. These points belong to the areas where units B, C, and D shift their active power consumption to reactive power consumption. If adding ramp constraints or forbidding such shifts, the feasible operating area will also reduce and become nonconvex.

Ignoring these effects of nonlinearities and nonconvexities can lead to overestimating the aggregated network flexibility and developing inaccurate models that are not applicable to real case studies. As demonstrated by the simulations, feasible network operating areas can be nonlinear and nonconvex when network constraints and additional limitations are imposed.
Such effects can be noticed even in existing studies. For example, in \cite{Silva2018}, even though not discussed by the authors, the flexibility areas constrained by the maximum flexibility costs are nonconvex (Fig.~2 in \cite{Silva2018}). It is important for future research to investigate the conditions under which the flexibility area remains convex. There is also a need to study possible coordination issues related to the power swap between flexible units and the rapid nonlinear shifts in unit operation.

\section{Conclusion}
This letter raises concerns about the nonlinearities and nonconvexities associated with ADN flexibility. Estimation of aggregated flexibility as a boundary of the feasible network operating area (or even approximating it by the Minkowski addition) overlooks the actions that units take to reach this boundary. As demonstrated in this work, due to the nonlinearity of the network power flow model, operation of flexible units exhibits highly nonlinear behavior, which requires perfect unit coordination and high ramping requirements.
If these effects of nonlinearities and nonconvexities are ignored, existing approaches in the literature can result in overestimation of ADN flexibility and give rise to impractical solutions.
Future research on the flexibility of distribution networks should accurately consider these effects when modelling and analysing operating points both at the boundary of the network feasibility set and within the boundary.


%





\ifCLASSOPTIONcaptionsoff
  \newpage
\fi



%




\bibliography{references.bib}
\bibliographystyle{IEEEtran}

%








\end{document}

%% file: packages.tex
\usepackage{amsmath, amssymb, amsfonts, amsthm, booktabs, cancel, comment, enumitem, eurosym, float, graphicx, mathtools, multirow, nicefrac, placeins, soul, steinmetz, stfloats, siunitx, svg, url, xcolor, tabularx}%
\usepackage{subfigure}
\usepackage{commath}
\usepackage{cite}
\usepackage{graphicx}
\usepackage{hyperref}
\usepackage{nomencl}
\usepackage{cleveref}
\makenomenclature

\floatstyle{ruled}
\newfloat{model}{thp}{lop}
\floatname{model}{\textsc{Model}}

\makeatletter
\renewcommand*{\fnum@model}{\fname@model}
\makeatother

\usepackage{breqn}

\usepackage{accents}